\documentclass[article, 12pt, twocolumn]{aastex631}
\usepackage{graphicx}
\usepackage{natbib}
\usepackage[utf8]{inputenc}
\UseRawInputEncoding
\begin{document}
\title{\bf Galactic kinematics of various stellar types}
\author{Charles R. Cowley}
\affiliation{Department of Astronomy, University of Michigan, 
1085 S. University, Ann Arbor, MI 48109-1107\\
orcid:0000-0001-9837-3662 \\
e-mail: cowley@umich.edu }                            
\author{Robert E. Stencel}
\altaffiliation{Department Astronomy, Univ. Michigan, Ann Arbor}
\affiliation{Chamberlin Observatory, University of Denver, 2930 E Warren Ave., 
     Denver, CO 80210, USA, \\
     orcid:0000-0001-8217-9435\\
     e-mails: robert.stencel@du.edu \\
     restence@umich.edu}


\begin{abstract}
We compare and contrast the kinematics of various classes of 
stars, both normal (with MK types) and peculiar. Numerous 
plots show the differences in spatial and velocity 
distributions as well as distributions in kinetic energy vs.
angular momentum ($KE\,vs.\,L_Z$) space. 
In the latter plots, young,  thin disk stars on nearly circular orbits cling to the left edge of a ``solar parabola" while older  objects on elliptical orbits fill the central parabolic region.   Some of the young 
Wolf-Rayet stars violate this trend due to smaller
semi-major axes than the Sun or orbital eccentricities.
Deviation  of the vertex of the velocity ellipsoid is discussed as an 
indication of population and youth, with an emphasis on Ap, 
Bp, Am, Fm, Herbig AeBe, and $\lambda$ Bootis stars. Both the vertex 
deviation and phase space distribution provide useful insights.
\end{abstract}

\section{Introduction}
In this paper we examine the kinematical properties of groups of stars
that are distinguished by their spectral peculiarities. 
Data in the Simbad and Gaia data bases \footnote{See Sec. Data for detailed
references.}  allow calculation in physical ($X$, $Y$, $Z$) and 
velocity space ($U$, $V$,$W$) with respect to the local standard of rest (LSR), here assumed to be 8.5, 13.58, and 6.49 km s$^{-1}$.  Numerical values for the latter 
were taken from PyAstronomy
\footnote{Code site: https://tinyurl.com/zkwzvrt8  or https://github.com/sczesla/PyAstronomy}
The present work uses a right-handed coordinate system in which $X$ and $U$ point 
to the Galactic center, while $Y$ and $V$ are in the direction of Galactic rotation.

We show plots of $Y$ vs. $X$, distributions in the Galactic plane for various categories of stars.  The present
work emphasizes different peculiar types rather than subclasses
of spectral types (e.g. G0 V, G2 V, etc.).   
The area covered by the stars generally reflect the intrinsic brightness
of the sample, such as O-stars vs G-dwarfs, and the area of the sky covered 
by the relevant survey.
The plots may show associations with young clusters or spiral arms, or
be associated with the sky fraction covered and/or the nature of the program from which the star sample was obtained.  In the case of G-dwarf solar twins, the spatial plots are of the solar neighborhood because the relevant survey was a search for extrasolar planets, best studied in nearby objects \citep{bd18}[henceforth BD18].

Plots of $V$ vs. $U$ for O-, Wolf-Rayet (henceforth, WR), and 
Herbig AeBe are dynamically unrelaxed, and show the influence
of clusters and spiral arms.  For later spectral types, A, B, and especially F and G,
the $U$ vs. $V$ plots are more ordered (relaxed) and  
may be considered to be a projection of a velocity ellipsoid.
This concept was introduced by \citet{ks07}, and widely
investigated over the years.  

A fine introduction of the velocity ellipsoid and
the related phenomenon of {\em vertex deviation} is in \citet{miro68}.  
Since the mid 20th Century, vertex deviation has been discussed in relation
to stellar populations \citep{vys57}.  The {\em angle} of deviation is measured
clockwise from the direction of Galactic rotation, and {\it tends to decline with
increasing age} \citep{bime98} -- see p. 631.  \citet{gom90} 
showed that the kinematics and vertex deviation of the 
local Population I dwarf stars (B5 V-F5 V), were related to age
\citep{sun23}. 
We note vertex deviations in Sec. ApBp-stars, AmFm-stars and lamBoo-stars.
However, attempts to measure relative ages from the angles of deviation have
proven unsatisfactory.  Some unweighted least squares fits to the points in UV space
give angles that are not consistent with eyeball fits, probably because of outliers.
We therefore show the relevant plots, and leave further quantitative evaluations
to future work.

We also show plots in $KE\,vs.\,L_Z$ space, where the coordinates are stellar kinetic energies ($KE$) and the z-component of the angular momentum $L_Z$, both 
defined per appropriate unit mass.  
For the plots, we take the origin of the
coordinate system to be the center of the Galaxy.  For these 
coordinates it is only necessary to substitute $X = X-R_c$, for the 
X-coordinate, and $V = V + V_c$, where $V_c$ is the circular velocity 
at the Local Standard of Rest (LSR).  We use 8 kpc for {\bf $R_c$}.

In the $KE\,vs.\,L_Z$
space plots, the solar $X$ is -8 kpc.
If we assume all of the kinetic energy is entirely due to 
a the circular velocity $V_c$ km/s, then 
\begin{equation}
KE =  \frac{1}{2}\, V_c^2.
\end{equation}  
For a circular orbit, the $Z$-component
of the angular momentum, $L_z = XV-YU$ and $L_z = R_c\times V_c$. 
It is common in calculations of this type to assume unit mass. 
 
From these simple relations we obtain a parabola
\begin{equation} 
KE = \frac{1}{(2\cdot R_c^2)}\times L_Z^2.
\label{eq:one}
\end{equation}
 This defines the {\it solar parabola} to which we shall henceforth refer.
Note that our $KE$ vs. $L_Z$ plot is qualitatively different from energy vs.
$L_Z$ plots where the energy is the total, kinetic plus potential
energy.  The Galactic potential energy may dominate the kinetic, so a plot of total energy vs. $L_Z$ {\it can be} qualitatively
different from one where only the kinetic energy is used.  We also neglect any
influence of the Galactic bar \citep{mor15}.

The Solar twins of BD18 
and \citet{spin18} are often used here as representative, thin disk stars.  
Indeed, most of our 
stars are thin-disk Population I and generally confined to the Galactic 
plane.  We show this explicitly  for the WR 
and Herbig AeBe stars,
but in the summary section (below), we give the average absolute value of the coordinate $Z$ for all of the data sets.

The Ap, Bp, Fp, HgMn, and $\lambda$ Bootis stars  
are subclasses of CP or chemically peculiar
stars of the upper main sequence \citep{pre74} and \citep{grco09}.  Their peculiar compositions are currently thought to originate primarily from endogenous elemental separations due to diffusion under radiation pressure \citep{mic70} with possible mass loss.  
A comprehensive review is by \citet{mar15}.  

\subsection{Radial velocities and parallaxes\label{sec:plxrv}}
Many radial velocities are missing from the Gaia DR3 for the O- and WR stars
 and those that are present are typically unreliable. 
Useful lines are often broader than in later types, and often partially
or wholly in emission.  In some of the more distant stars, there can be
perturbations from interstellar  material.  For that reason for O- and 
WR stars we have used estimates based on the transverse velocities
in right ascension and declination; 
We assumed that the radial velocity 
is equal to the mean of the two transverse velocities, with alternating
positive and negative values.  We designate these as ``the {\it estimated} RVs".
We note but have not implemented statistical ways of estimating the
radial velocities \citep{nai23}, which require knowledge of stars near 
the O- or Wolf Rayet star.  It is beyond the scope of the present work
to obtain such data.  As an alternative, we have calculated the distributions
for A and F stars (which have good radial velocities) in UV space using both 
the spectroscopic and estimated radial
velocities.  The general distribution of points in both cases were
quite similar.  

Parallax errors are of primary concern. 
Errors of proper motion may make minor distortions in our plots but do not 
lead to serious mistakes such as giving unrealistic distances.
We follow 
Plotnikova et al. (2023) who considered four different methods 
obtaining the best parallaxes from Gaia, and adopted those coming 
directly from Gaia DR3.   

We considered a sample of Gaia DR3 Galactic O-Star Catalogue
(GOSC) of \citet{mawal04} O-stars with small 
parallaxes $\le 0.2$ milliarcsec (mas).  While there was considerable
scatter, the parallax errors averaged roughly 0.1 times the the
parallaxes.  
We also considered Gaia DR3 parallax over error vs. parallax. 
Again, there was 
is considerable scatter, showing a parallax over error value of 5  
was obtained for parallax values of 0.08 to 0.1  mas. We 
have chosen the more conservative value and excluded all stars 
with parallax values less than 0.1 mas for all sets of 
stars in the present study.


For the WR stars we used distances estimated by P. Crowther (private communication) and considered ``good".  We did not consider his higher and lower distance estimates. Bayesian re-considerations of this point are not likely to improve on these estimates until alternate methods of radial velocity determination are found.

\section{The kinematics of groups of stars}

\subsection{O-stars}
\label{sec:O-stars}

\begin{figure*}[!t] \centering
\includegraphics[width=\textwidth]{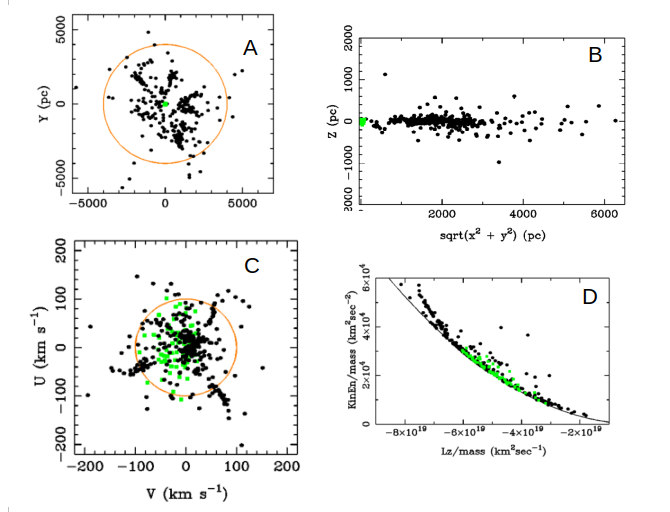}    
\label{fig:O-stars}
\caption{
A. O star distribution in physical space from the GOSC Star Catalogue (Maiz Apellaniz et al. 2004). 
The G-dwarf, BD18 solar twins (merged green squares) are at the center of the plot and present a strong contrast. 
The orange circle is 4 kpc in radius. 
The G dwarf, BD18 solar twins (merged green squares) are at the center of the plot and present a strong contrast.  
The orange circle is 4 kpc in radius. 
B. Z-axis distribution. 
C. The distribution of (GOSC) Ostars in velocity space (black filled circles) is only a little greater than that of of BD18 solar twins in green squares, in contrast to the X −Y plot for physical space. 
Both sets of stars are in or near the 100 km/s orange circle. 
D. O stars of all types (black filled circles) from the GOSC catalogue are plotted in KE vs LZ space.
Compare these with the BD18 sample.}
\end{figure*} 

We discuss 354 O-stars of all subclasses from the Galactic O-Star Catalogue
(GOSC) of \citet{mawal04}.
Fig. 1A,B show that the GOSC O-stars are distributed far more
widely in space than the solar twins of BD18 (merged green squares) which are all
contained within a radius of about 100 pc.  This is expected, as the BD18
twins contain only dwarfs chosen in planet searches, 
while the GOSC O-stars contain giants as well
as dwarfs.  

In (UV) velocity space (Fig.1C), both the O-star
and solar twin motions fall mostly within a circle with 100 km/s
radius shown.  The O-stars show grouping in both physical and
velocity space suggestive of clusters or spiral arms.  
The $U$ and $V$ points would be placed somewhat differently if actual
rather than estimated radial velocities were used.

For the $KE\,vs.\,L_Z$ space plot (Fig.1D) we again used radial velocity estimates based on the
tangenial motions.  
Because our $X$ is zero for the center of the Galaxy and negative 8 kpc
for the solar neighborhood, $L_z$ is a large negative number for stars
on prograde orbits (positive $V$). 
The BD18 solar twins are plotted as green squares
while the GOSC O-stars are black, filled circles.  Generally, young stellar
populations cling to the parabola as do most of the solar twins.  

Stars may fall into the interior of the solar parabola
for various reasons. They may be on elliptical orbits or have
circular orbits with radii less than that of the Sun.  Very
metal poor stars typically have elliptical and sometimes 
retrograde orbits as they belong to an old population.  

Two high outliers
near the center of the $KE\,vs.\,L_Z$ plot are HD 157857, an O6.5 II(f) (upper left), and
CD -26 5136 (lower right), a Blue Supergiant O 6.5 Iabf, according to
Simbad.

\subsection{Wolf-Rayet (WR) stars}

\begin{figure*}[!t] \centering
\includegraphics[width=\textwidth,height=12cm]{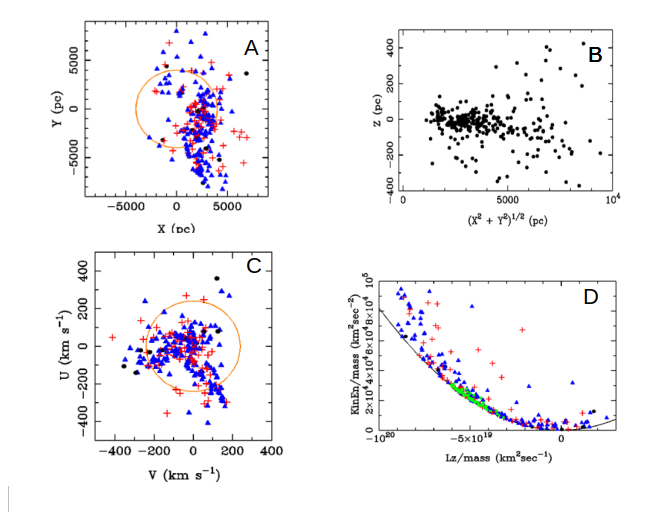}    
\caption{A. X and Y positions in the Galactic plane of WN
(blue triangles) and WC (red plus signs) stars from CCat. The N- and C-types are intermixed. WR stars that are neither ‘N’ nor ‘C’ are plotted as black filled circles. The orange circle is 4 kpc in radius.  B. WR stars from CCat showing the vertical (Z) direction. The envelope is similar to the ± 250 pc discussed by Rustamov \& Abdulkarimova (2023).  C. WR stars from CCat in velocity space. Symbols and colors are as before. The orange circle is at 240 km/s.  D. KE vs. LZ plot of WN (blue triangles) and WC
(red plus signs) stars. BD18 twins are green squares.}
\label{fig:WR-stars}
\end{figure*}

The WR stars are a heterogeneous group of evolved massive
stars.  Most show strong emission lines of either nitrogen (WN stars) or carbon
(WC stars).  For a more detailed description of WR types see \citet{cro07}.
Fig.~\ref{fig:WR-stars}A shows the distribution in physical space of
WR stars from the catalogue
of Paul Crowther and colleagues \citep{rc15a}, henceforth CCat 
(see also \citet{rc15b,cro07}). 
\footnote{https://pacrowther.staff.shef.ac.uk/WRcat/}
 
The distances are from a private communication kindly sent by Paul Crowther.  Only 
values which he considered good estimates were plotted.  
We used a crude 
classification where the red plus signs are from stars with
the letter `C' in the CCat classification, while blue triangles are for stars with `N'. 
Neither subtype shows a clear preference for a particular area of physical
space. 

The Galactic WR stars
are clumped at a distance from the Galactic center that is
somewhat less than that of the Sun, and they 
spread in the Y-direction--that of Galactic rotation.
Fig.~\ref{fig:WR-stars}A may be compared with the plot in a recent paper by 
\citet{nm23} giving positions of N- and C-subtype WR stars in the Andromeda
galaxy.  Their Fig. 1 shows an elliptical ring of N and C\, WR types in
about the same ratio as in our sample:  C/N = 0.70, 0.63 (Galaxy, Andromeda). 
The WR stars in our Fig.~\ref{fig:WR-stars}A might represent a fraction of a 
similar ring in our own Galaxy. 

Our results for the distribution of WR stars in (Galactic) physical space
is very similar to those of \citet{rusabd23}, who used a somewhat larger
sample of WR stars.

Fig.~\ref{fig:WR-stars}C shows the WR stars in velocity space. 
The dispersion in velocities of the WR stars is significantly larger then
that of the solar twins which would fall mostly within a 63 km/s circle.
(see Fig.~\ref{fig:WR-stars}C)  
The WC and WN stars are intermixed.  
The magnitude and spread of the WR
velocities is surprisingly large for a population of young stars.  We note the
significant, non-symmetric structure of the velocities, details of which
are affected by the radial velocity approximations.

Fig.~\ref{fig:WR-stars}B shows the distribution of WR stars in the vertical
direction and shows the concentration to the Galactic plane with the southern
hemisphere somewhat more populated.  The vertical envelope is
similar to the $\pm$ 250 pc discussed by \citet{rusabd23}.  If we
assume a lifetime of 1 Myr for a WR star  
at 240 km/s (see Fig.~\ref{fig:WR-stars}C), the star could travel 250 pc
in about 1 Myr, (using 1 km/s = 1 pc in $10^6$ years).  This is the same
order of magnitude as the estimate of a representative WR 
evolutionary phase \citep{rc15a}.

A plot of
$KE\ vs.\ L_Z$ (Fig.~\ref{fig:WR-stars}D) gives additional perspective.
A sizable fraction of the stars cling
to or are near the negative branch of the solar parabola, where young stars reside.
There is some scatter, even to positive $L_Z$, implying retrograde
orbits.  Such orbits are typical of some very old, metal-poor dwarfs
 \citep{mor15,cs23} and Table 1.  Binary
ejection may be responsible for some of the unusual velocities, which are
difficult to understand for young WR stars.  Some of the scatter to higher
$L_Z$ and $KE$ is surely due to radial velocity estimates.  
The WR stars with the largest
(positive) values of $L_Z$ are: WR 29, WR 30a, and WR 61.
We find no indication from Simbad or CCat that would explain their unusual
$L_z$ values.  

An older catalogue by
\citet{vdh01} shows some WR stars lifting off the parabola, but shows no
points in the positive (retrograde) $L_Z$ side.

\subsection{Mercury-Manganese (HgMn) stars}	
\begin{figure*}[!t]
\includegraphics[width=\textwidth, height=12cm]{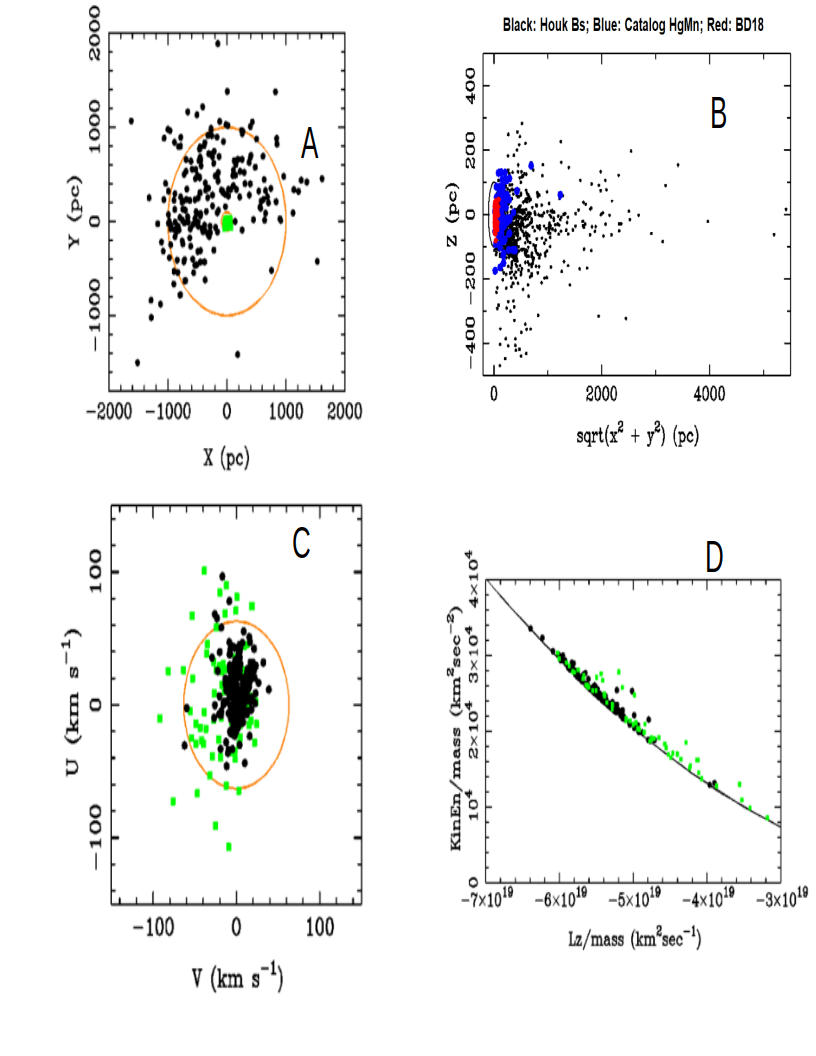}    
\label{fig:HgMn-stars}
\caption{A. HgMn stars in physical space from Chojnowski et al. (2020) (black filled circles). The BD18 solar twins green squares) are shown for perspective. The inner orange circle has a radius of 100 pc, the outer circle 1000 pc. The distribution is asymmetrical because of the sky survey coverage.  
B. Z axis locations with respect to galactocentric distance.  
C. HgMn stars in velocity space from Chojnowski et al. (2020), i.e. black filled circles. The BD18 solar twins (green squares) are shown for comparison. The circle has a radius of 63 km/s.  
D.  HgMn stars in KE vs LZ space (black filled circles). The green squares are BD18 solar twins.}
\label{fig:HgMn-stars}
\end{figure*}

HgMn stars are chemically peculiar, mostly B dwarfs, with weak or no measurable magnetic fields.      
\citet{ghal16} describe the properties of this class and present
a catalogue of abundances for 81 of these stars.  They can have unbelievably
bizarre abundances of heavy elements, such as a million times solar for mercury.
A few of these stars show puzzling isotopic excesses of the rare $^{48}$Ca
\citep{cahu04}.  
While the HgMn stars may also be
considered Bp, only one of the 81 stars are in the set of ApBp stars
of Section ~\ref{sec:ApBp-stars}.  This is because the HgMn types were nearly
impossible to detect at the resolution of the Houk-Michigan 
\citep{hosw00} spectra.  We
therefore discuss them as a separate class.
Fig.~\ref{fig:HgMn-stars}AB  shows the distribution in physical space of the 264
stars from \citet{chohub20} survey.   

In Fig.~\ref{fig:HgMn-stars}C, we plot the
\citet{chohub20} stars in velocity space.  They are nearly all contained within
the 63 km/s circle of low velocity stars.  
\footnote{Near the middle of the 20th century, velocities in the
 mid 60's (km/s) were cited by \citet{oo30} as a limit between low and high velocity
 stars--6 km/s became a favorite value.  See also \citet{sc60}} Indeed, their scatter is 
tighter than that of the BD18 solar twins.  This should be contrasted with
the wide distribution of the HgMn stars in physical space ($XY$) where
HgMn stars scatter beyond one kpc.  
This broad scatter, surely, is only partially
due to the great depth of the \citet{chohub20} survey.
The BD18 stars are all within 100 pc of the sun. 
  
These HgMn stars are shown in Fig.~\ref{fig:HgMn-stars}D in $KE\,vs.\,L_Z$ space. 
One of the modest outliers, 38 Dra, is noted in Simbad as a high proper motion
star.  Another, HD 53004 is described in Simbad as an eclipsing binary.

\subsection{Herbig AeBe stars}
\begin{figure*}[!t]
\includegraphics[width=\textwidth, height=12cm]{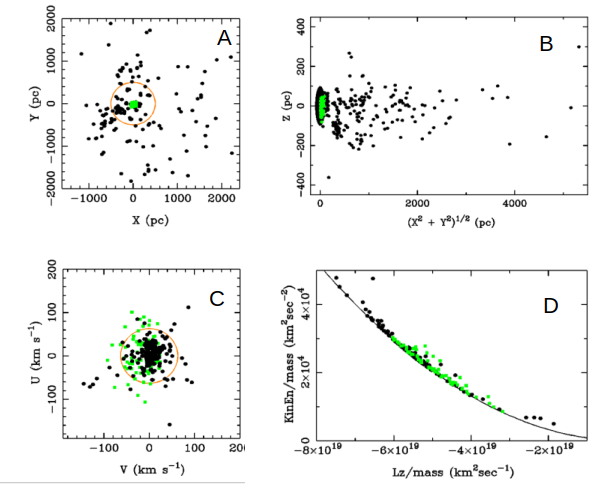}

\caption{ A. Herbig AeBe stars in physical space. The filled
black circles are the Herbig AeBe stars from Vioque et al.
(2018), the green squares within the orange circle of 500 pc
are BD18 solar twins.  
B.  The vertical extent of the Herbig AeBe stars
(extra symbols as annotated). 
C. Herbig AeBe stars in velocity space (black). The
circle is at 63 km/s . Unlike the spatial distribution, the velocities of the Herbig AeBe stars are
similar to those of the solar twins.
D.  Herbig AeBe (black) and BD18 (green) stars
in KE vs. LZ space. The plot is similar to other KE vs. LZ
plots for young objects. }
\label{fig:HerbigAeBe-stars}
\end{figure*}

We examine the kinematics of 215 Herbig AeBe stars from the study by \citet{vio18}.   
Herbig stars \citep{wawa98}[henceforth, Herbigs] may be considered more massive (2-10$M_\odot$) analogues of the pre-main sequence T-Tauri stars.  They are distinguished from the Ap, Am, and Bp types discussed below by their emission lines and infrared excesses which originate from their star-forming envelopes.   They have a wide distribution in physical space as shown by Fig.~\ref{fig:HerbigAeBe-stars}AB with a clear preference for the direction of the Galactic center (positive X).
 
Among the \citet{vio18} Herbig stars only about 30\% have radial velocities in Gaia, so
the remaining were estimated from the transverse velocities as was done for the 
O- and WR stars.  Again the $UV$ values are only approximate for those stars,
but the overall character of the plots are not changed by the radial velocity
estimates.
The velocity space plot is shown in Fig.~\ref{fig:HerbigAeBe-stars}C.  The distribution of 
velocities is similar to the BD18 solar twins, as was the case for the O-stars
of Fig.~\ref{fig:O-stars}.

In $KE\,vs.\,L_Z$ space (Fig.~\ref{fig:HerbigAeBe-stars}D) the Herbigs cling tightly to the left wing of the parabola, as would be expected for a very young population.

\subsection{Ap, Bp types and superficially normal A and B dwarfs}
\label{sec:ApBp-stars}

\begin{figure*}[!t]
\includegraphics[width=\textwidth, height=8cm]{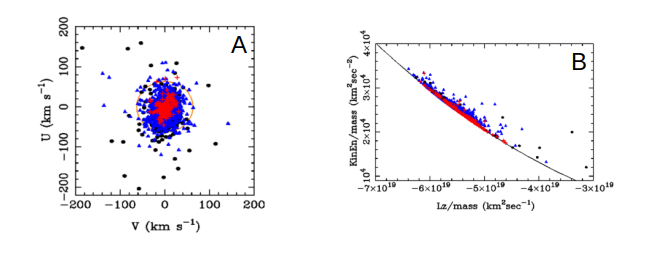}
\caption{ A. Normal dwarfs: A stars (Blue triangles)
and B stars (Black filled circles) and peculiar types,
ApBp (red crosses) from Houk-Michigan catalogues.
The latter form a compact group, almost entirely within the
63 km/s circle, and showing a vertex deviation indicating
young objects. B. A striking feature of this KE − LZ diagram
is how closely the ApBp stars (red, crosses) cling to the
parabola boundary. A few A dwarfs (blue triangles)
scatter away from the boundary, and some B dwarfs 
(black filled circles) scatter somewhat more.}
\label{fig:ApBp-stars}
\end{figure*}

Stars that are not specifically designated as peculiar have MK types assigned
in the  Houk-Michigan catalogues. 
We take them to be `superficially normal,' or simply
`normal.'\footnote{Stellar spectroscopists have said that all stars
are chemically peculiar if examined at high enough dispersion.} 
Plots in X vs Y of data from \citet{hosw00} are irregular in shape
because of the partial sky coverage of the catalogues and are not shown.  
In velocity space, Fig.~\ref{fig:ApBp-stars}A,
the points are distributed approximately circularly for A and B dwarfs.  The
ApBp types show a vertex deviation. 
The B dwarfs are largely obscured
by the slightly wider A-dwarf distribution.  The red points for ApBp stars show 
a vertex deviation.
 
These same stars are shown in the $KE\,vs.\,L_Z$ space 
of Fig.~\ref{fig:ApBp-stars}B.

Five of the B-star outliers in Fig.~\ref{fig:ApBp-stars}B have some abnormalties: three are called {\it high proper motion}, another a double 
or multiple, and a fifth a hot subdwarf by Simbad.

The behavior of the (red) crosses for the Ap and Bp stars is as expected if we
attribute their peculiarities to endogenous diffusion 
and assume those effects diminish with age \citep{blb14}.
These young stars appear close to the $KE-L_Z$ parabola, and hardly stray 
into the region where older, metal-poor stars are found.  The vertex 
deviation of Fig.~\ref{fig:ApBp-stars}A could indicate relative youth.

\subsection{Am and Fm stars and normal A and F dwarfs from Houk-Michigan Catalogues}
\label{sec:AmFm-stars}

The physical space plot of F dwarfs and AmFm stars is also not shown 
because of the partial coverage of the Houk-Michigan catalogues.  
The velocity space plot, Fig.6 
is similar to that of 
Fig.~\ref{fig:ApBp-stars}A, though slightly more compact.  
Note that the normal A dwarfs (blue triangles) are
not plotted.  The AmFm stars show a vertex deviation.  

\begin{figure*}[!t]
\includegraphics[width=\columnwidth]{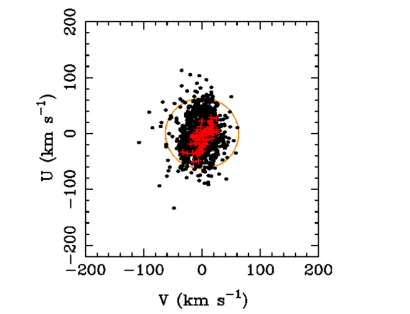}
\caption{Similar to Fig.~\ref{fig:ApBp-stars}A but for normal
 F dwarfs (black filled circles), and AmFm stars (red crosses).
 The circle is 63 km/s.}
\label{fig:AmFm-stars}
\end{figure*}

\subsection{$\lambda$ Bootis stars}
\label{sec:lamBoo-stars}
\begin{figure*}[!t]
\includegraphics[width=\textwidth, height=13cm]{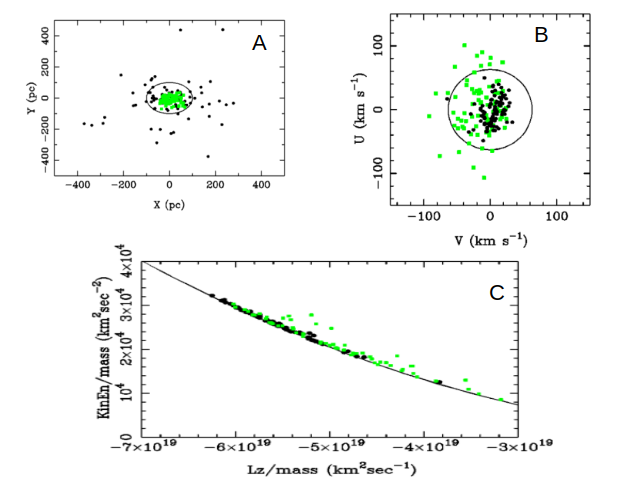}
\caption{ A. The black filled circles are $\lambda$ Bootis stars, while
the green squares are BD18 solar twins. B. $\lambda$ Bootis stars in velocity space. The black points arguably show a vertex deviation.  C. $\lambda$ Bootis stars in KE vs. LZ space.}
\label{fig:lamBoo-stars}
\end{figure*}

\citet{grco02} describe the fascinating and chemically peculiar, upper main sequence stars,
named after the prototype $\lambda$ Bootis.  Their abundance pattern is a
depletion of elements with high condensation temperatures \citep{lod03} while elements
with low condensation temperatures have nearly solar abundances.  This pattern is widely seen in a variety of astronomical settings from some Herbig AeBe stars \citep{fbw12} and the interstellar medium \citep{vl90}. 

The stars in the following plots were taken from \citet{mur15} who provided
a list of 212 possible $\lambda$ Bootis stars.  Only stars 
with ``LamBoo" in the list were used here resulting in 90 stars. 
The dispersion in physical space of the $\lambda$ Bootis stars \ref{fig:lamBoo-stars}A is noticeably larger than
that of the BD18 solar twins
because of their higher luminosities.  
In velocity space \ref{fig:lamBoo-stars}B, the relation is reversed, with the 
$\lambda$ Boo stars more nearly equal to the Sun's velocity. 
 
In $KE\,vs.\,L_Z$ space, \ref{fig:lamBoo-stars}C, the $\lambda$ Bootis stars are as closely crowded to the
parabola as any stars of our samples.    

\subsection{G-stars}
We have used the BD18 solar twin stars as typical G-stars.  However, these 79 stars
were chosen from a survey for possible planet hosts \citep{ram14}.  It is therefore
useful to compare their properties with the somewhat more numerous (222) G dwarfs of the
BSC \citep{bsc5}.  Kinematic plots are not shown, but Table 1
presents numerical values that show the BD18 and BSC G-stars
have quite similar kinematic properties.

\section{Summary}
\label{sec:sum}

The spatial (XY) distributions of Ap, Bp, Am, Fm, and $\lambda$ Bootis
stars fall mostly within a several hundred pc radius of the Sun.The distributions
depend on sky coverage and the depth of the relevant surveys as well as the 
intrinsic properties of the populations.  This is expecially clear in the XY plot
for the O-stars, where the points scatter beyond 4 kpc because of their high luminosities.  
The XY plot of the WR stars extends beyond 5 kpc, and has an irregular structure.

The velocity (UV) distributions of the Ap, Bp, Am, Fm, and $\lambda$ Bootis
stars are all similar to chemically normal stars near the Sun.  In the
$KE\,vs.\,L_Z$ plots, these types show a proclivity to fall on the solar
parabola.  The O-stars show some scatter off the parabola, while the WR
types show puzzling $L_Z$ values including retrograde orbits.
A summary of relevant numerical values for the various sets of
stars is given in the following table, below.

\begin{deluxetable*}{lccccc}  [!ht] 
\tablecaption{Stellar Data Sets--Number of stars, data source, mean value of $|Z|$, average KE per unit mass, 
mean dist from left branch of parabola and standard deviations\label{tab:1}.  Ultra metal-poor stars
are added \citep{cs23} to emphasize their extreme $KE\,vs.\,L_Z$ space distributions due to high
orbital eccentricity and/or lower circular velocities.}
\tablecolumns{6}
\tablewidth{900pt}
\tablehead{
\colhead{Data Set} &           
\colhead{No. Stars} &           
\colhead{Source} &             
\colhead{Mean$|Z|$(pc)}&       
\colhead{10$^{-3}KE[(km\ s^{-1})]^2$}& 
\colhead{$10^{-18}$ mean dist}    
}
\startdata                           
O-stars    & 354  & \citet{mawal04} &$84\pm 123  $  &$27.9 \pm 15.7$&$ 2.21\pm 4.85 $   \\
B-stars    &  62  & \citet{hosw00}  &$97\pm 133 $  &$24.1 \pm 4.16$&$ 0.86\pm 2.30 $   \\
A-stars    &2138  & "               &$128\pm 111$  &$25.9 \pm 2.61$&$ 0.34\pm 0.50 $   \\
F-stars    &2718  & "               &$90\pm 85  $  &$24.7 \pm 4.09$&$ 0.77\pm 3.54 $   \\  
G-stars    & 222  & \citet{bsc5}    &$24\pm 32  $  &$27.9 \pm 0.29$&$ 0.36\pm 2.55 $    \\
BD18       &  79  & \citet{bd18}    & $23\pm  17$  &$22.2 \pm 5.13$&$1.28 \pm 1.56 $  \\
Wolf-Rayet & 274  & \citet{rc15a}   &$-30.6\pm 144$  &$34.6 \pm 30.8$&$10.4 \pm 1.5 $\\ 
ApBp       & 143  & \citet{hosw00}  &$62\pm 59  $  &$24.7 \pm 2.88$&$0.27 \pm  0.43$    \\   
AmFm       &  79  & "               &$49\pm 36  $  &$31.6 \pm 1.22$&$1.17 \pm  2.71$   \\
HgMn       &  81  &\citet{chohub20} &$62\pm 42  $  &$24.1 \pm 0.27$&$0.25 \pm  0.71$   \\
Herbig AeBe& 215  & \citet{vio18}   &$78\pm 67  $  &$25.7 \pm 6.52$&$0.86 \pm  4.68$   \\ 
$\lambda$ Boo&90  &\citet{mur15}    &$70\pm 64  $  &$25.4 \pm 3.21$&$0.31 \pm  0.30$    \\
UMetalPoor & 299  & \citet{ro14}    &$1447\pm 1703$&$42.48\pm 39.9$&$39.89\pm 0.00 $   \\
UMetalPoor &246   & \citet{bar05}   &$7590\pm 13990$&$55.4 \pm 50.9$&$50.89\pm 0.00 $ \\ \tableline
\enddata         
\end{deluxetable*}

\section{Data availability}
\label{sec:data}
The data underlying this article are available in the articles referenced and the 
Simbad and Gaia data bases.

\section{Acknowledgements}
\label{sec:ack}
We are grateful to  M. Valluri for clarification of various aspects of Galactic structure
and kinematics.  P. Crowther kindly provided an ASCII version of his table of 
WR stars along with distance estimates.  S. Hubrig and
M. Sch\"oller kindly sent us an ascii version of the \citet{chohub20} table of HgMn stars.
Thanks are due to J. Fernandez-Trincado and C. Reyl for help with the Galactic
potential energy. We note that an anonymous reviewer provided many practical comments. 
We thank D. Nidever and D. Chojnowski for advice on the reliability of APOGEE radial velocities
of WR stars.  Thanks also to S. Oey for pointing out ejection as a possible 
explanation for some extreme velocities of WR stars.  The second author is sad to note that the first author recently has been moved into hospice care.
We used PyAstronomy, \url{https://github.com/sczesla/PyAstronomy}.

This work has made use of data from the European Space Agency (ESA) mission
Gaia (\url{https://www.cosmos.esa.int/gaia}), processed by the Gaia
Data Processing and Analysis Consortium (DPAC,
\url{https://www.cosmos.esa.int/web/gaia/dpac/consortium}). Funding for the DPAC
has been provided by national institutions, in particular the institutions
participating in the Gaia Multilateral Agreement.
We also acknowledge use of the SIMBAD database \citep{wen00} operated at
CDS, Strasbourg, France.  We are grateful for the support of the Michigan
Astronomy Department and help from the LSA Technology Services.


\end{document}